\documentclass[aps,prb,twocolumn,showpacs,superscriptaddress]{revtex4}

\usepackage{graphicx}
\usepackage{amsmath}
\usepackage{amssymb}

\begin{document}

\title{Autocorrelation of quasiparticle spectral intensities and its connection with quasiparticle scattering interference in cuprate superconductors}

\author{Deheng Gao\footnotemark[1]}

\affiliation{Department of Physics and Optoelectronic Engineering, Weifang University, Weifang 261061, China}
\affiliation{Department of Physics, Beijing Normal University, Beijing 100875, China}

\author{Yingping Mou\footnotemark[1]}

\affiliation{Department of Physics, Beijing Normal University, Beijing 100875, China}

\author{Yiqun Liu}

\affiliation{Department of Physics, Beijing Normal University, Beijing 100875, China}

\author{Shuning Tan}

\affiliation{Department of Physics, Beijing Normal University, Beijing 100875, China}

\author{Shiping Feng\footnotetext[1]{These authors contributed equally to this work}\footnote[2]{E-mail address: spfeng@bnu.edu.cn}}

%\author{Deheng Gao\footnotemark[1], Yingping Mou\footnotemark[1], Yiqun Liu, Shuning Tan, and Shiping Feng\footnotetext[1]{These authors contributed %equally to this work}\footnote[2]{E-mail address: spfeng@bnu.edu.cn}}

\affiliation{Department of Physics, Beijing Normal University, Beijing 100875, China}

\begin{abstract}
The quasiparticle excitation is one of the most fundamental and ubiquitous physical observables in cuprate superconductors, carrying information about the bosonic glue forming electron pairs. Here the autocorrelation of the quasiparticle excitation spectral intensities in cuprate superconductors and its connection with the quasiparticle scattering interference are investigated based on the framework of the kinetic-energy driven superconducting mechanism by taking into account the pseudogap effect. It is shown that the {\it octet} scattering model of the quasiparticle scattering processes with the scattering wave vectors ${\bf q}_{i}$ connecting the hot spots on the constant energy contours is intrinsically related to the emergence of the highly anisotropic momentum-dependence of the pseudogap. Concomitantly, the sharp peaks in the autocorrelation of the quasiparticle excitation spectral intensities with the wave vectors ${\bf q}_{i}$ are directly correlated to the regions of the highest joint density of states. Moreover, the momentum-space structure of the autocorrelation patterns of the quasiparticle excitation spectral intensities is well consistent with the momentum-space structure of the quasiparticle scattering interference patterns observed from Fourier-transform scanning tunneling spectroscopy experiments. The theory therefore confirms an intimate connection between the angle-resolved photoemission spectroscopy autocorrelation and quasiparticle scattering interference in cuprate superconductors.
\end{abstract}


\maketitle

\section{Introduction}

The nature of the quasiparticle excitations in cuprate superconductors is of great interest in the past three decades \cite{Zhou18,Damascelli03,Campuzano04,Kordyuk10,Devereaux07,Fischer07}. This follows an experimental fact that the parent compound of cuprate superconductors is a strongly correlated Mott insulator, which is realized by the localization of an electron at each copper atom of the copper-oxide planes in real-space \cite{Bednorz86,Kastner98}. However, when a small fraction of these electrons are removed from the copper-oxide planes, a process so-called charge-carrier doping, the electronic correlations are altered sufficiently to produce superconductivity, which is characterized by the delocalization of the electron pairs \cite{Zhou18,Damascelli03,Campuzano04,Kordyuk10,Devereaux07,Fischer07,Bednorz86,Kastner98}. This remarkable evolution from the localized real-space state of the Mott insulator to the delocalized momentum-space electron pairs of the superconductor therefore leads to a rich phenomenology in cuprate superconductors \cite{Timusk99,Hufner08,Comin16,Anderson87,Phillips10}. In particular, since the notable properties of the electronic state are intimately connected to the particular characteristics of the low-energy quasiparticle excitations \cite{Zhou18,Damascelli03,Campuzano04,Kordyuk10,Devereaux07,Fischer07}, the understanding of the nature of the quasiparticle excitations in cuprate superconductors is thought to be key to the understanding of how a strongly correlated Mott insulator with the localized electronic state becomes a superconductor with the delocalized electron pairing-state.

Experimentally, angle-resolved photoemission spectroscopy (ARPES), which probes the energy and momentum of electrons simultaneously, is a direct tool in the measurement of the momentum-space electronic structure of the system \cite{Zhou18,Damascelli03,Campuzano04,Kordyuk10}. In particular, the ARPES experimental measurements have obtained rather detailed information of the quasiparticle excitations of cuprate superconductors in the pseudogap phase \cite{Zhou18,Damascelli03,Campuzano04,Kordyuk10,Norman98,Yang08,Meng09,Yang11,Chan16}, where one of the most definite characteristics is the electron Fermi surface (EFS) reconstruction \cite{Norman98,Yang08,Meng09,Yang11,Chan16}, i.e., although the quasiparticle excitations of cuprate superconductors in the SC-state are well defined at all momenta along EFS, the weight of the quasiparticle excitation spectrum around the antinodal regime is gapped out by the pseudogap, and then EFS is broken up into the disconnected Fermi pockets located around the nodal regime. The Fermi pocket is consisted by the Fermi arc and back side of Fermi pocket. However, the highest intensity points are located at the tips of the Fermi arcs \cite{Chatterjee06,Shi08,Sassa11,Comin14,Horio16}, and then these tips of the Fermi arcs connected by the scattering wave vector ${\bf q}_{i}$ contribute effectively to the quasiparticle scattering processes \cite{Comin16,Chatterjee06,Comin14,Horio16}, which lead to the unconventional electronic state properties in cuprate superconductors. Moreover, this pseudogap also induces a dramatic change of the quasiparticle excitation spectral line-shape, where a sharp peak develops at the lowest binding energy corresponding to the superconducting (SC) gap, and is followed by a dip and then a hump in the higher energies, giving rise to a striking peak-dip-hump (PDH) structure in the quasiparticle excitation spectrum \cite{DLFeng02,Borisenko03,DMou17}. On the other hand, scanning tunneling spectroscopy (STS) is a direct tool in the detection of the real-space inhomogeneous electronic structure of the system \cite{Devereaux07,Fischer07}. In particular, this STS measurement technique has been also used to infer the momentum-space behavior of the quasiparticle excitations of cuprate superconductors in the pseudogap phase from the Fourier transform (FT) of the position- and energy-dependent local density of states (LDOS) $\rho({\bf r},\omega)$, therefore both real-space and momentum-space modulations for LDOS in the pseudogap phase are explored simultaneously \cite{Devereaux07,Fischer07,Pan01,Kohsaka07,Hanaguri07,Kohsaka08,Hanaguri09,Kondo09,Vishik09}. The typical feature observed by the FT-STS LDOS $\rho({\bf q},\omega)$ is dominated by the sharp peaks at the well-defined wave vectors ${\bf q}_{i}$ obeying the {\it octet} model \cite{Devereaux07,Fischer07,Pan01,Kohsaka07,Hanaguri07,Kohsaka08,Hanaguri09,Kondo09,Vishik09}, since the quasiparticle dispersion has closed constant-energy Fermi pockets around nodal regime. The quasiparticle scattering interference (QSI) manifests itself as a spatial modulation of $\rho({\bf r},\omega)$ with these well-defined wave vector ${\bf q}_{i}$, appearing in the FT-STS LDOS $\rho({\bf q},\omega)$. Although the STS experiments also indicated that the intensity of some of the QSI peaks in cuprate superconductors vanishes beyond the antiferromagnetic (AF) zone boundary, it has been shown this extinction of QSI without implying the loss of the quasiparticle excitations beyond the AF zone boundary \cite{Vishik09}. Furthermore, the experimental observations from the ARPES measurements have shown that the sharp peaks of the ARPES autocorrelation are directly correlated with the wave vectors ${\bf q}_{i}$ that connect the tips of the Fermi arcs \cite{Chatterjee06}, and are well consistent with these observed from the FT-STS experiments \cite{Pan01,Kohsaka07,Hanaguri07,Kohsaka08,Hanaguri09,Kondo09,Vishik09}. In this case, a natural question is why there is a direct connection between the sharp peaks of the ARPES autocorrelation detected in the ARPES measurements and the QSI peaks observed from the FT-STS experiments?

Theoretically, the quasiparticle excitation spectrum of cuprate superconductors in the pseudogap phase and the unusual behavior of QSI have been studied extensively \cite{Zhou18,Damascelli03,Campuzano04,Kordyuk10,Devereaux07,Fischer07}. In particular, it has been shown that the sharp peak observed in the quasiparticle scattering rate of cuprate superconductors is directly responsible for the remarkable PDH structure in the quasiparticle excitation spectrum \cite{Gao18}. On the other hand, the qualitative behaviors of QSI in cuprate superconductors have been discussed based on the phenomenological {\it octet} model by considering the effect of impurity scattering \cite{Fischer07,Wang03,Wang10,Nunner06,Zhang03}, where it has been shown that a single or few impurities in a homogeneous d-wave SC-state leads to a result that is in qualitative agreement with the FT-STS experimental data \cite{Pan01,Kohsaka07,Hanaguri07,Kohsaka08,Hanaguri09,Kondo09,Vishik09}. However, to the best of our knowledge, the ARPES autocorrelation in the pseudogap phase of cuprate superconductors and its connection with QSI have not been discussed starting from a microscopic SC theory, and no explicit calculations of the energy dependence of the ARPES autocorrelation has been made so far. In this paper, we study this issue by taking into account the pseudogap effect. Within the framework of the kinetic-energy-driven SC mechanism \cite{Feng15,Feng0306,Feng12}, we evaluate explicitly the autocorrelation function of the quasiparticle excitation spectral intensities in cuprate superconductors in terms of the electron spectral function, and reproduce the main experimental results of the ARPES autocorrelation of cuprate superconductors \cite{Chatterjee06}. In particular, our results show that the highly anisotropic momentum-dependence of the pseudogap gaps out the quasiparticle excitation spectral weight on the constant energy contours around the antinodal region, leaving behind the quasiparticle excitation spectral weight located at the disconnected segments around the nodal region only to form the Fermi pockets, where the highest intensity regime on the disconnected segments appears exactly around the tips of these disconnected segments, which in this case coincide with the hot spots on the constant energy contours, and then the quasiparticle scattering processes with the scattering wave vectors ${\bf q}_{i}$ connecting the hot spots construct a {\it octet} scattering model. As a consequence, the sharp peaks in the autocorrelation of the quasiparticle excitation spectral intensities with the scattering wave vectors ${\bf q}_{i}$ are directly associated with the regions of the highest joint density of states. Furthermore, the momentum-space structure of the autocorrelation patterns of the quasiparticle excitation spectral intensities is well consistent with the momentum-space structure of the QSI patterns observed from the FT-STS experiments \cite{Pan01,Kohsaka07,Hanaguri07,Kohsaka08,Hanaguri09,Kondo09,Vishik09}.

This paper is organized as follows. The basic formalism is presented in Sec. \ref{Formalism}, while the quantitative characteristics of the ARPES autocorrelation of cuprate superconductors and its connection with QSI are discussed in Sec. \ref{ARPES-autocorrelation}, where we confirm an intrinsic connection between the sharp peaks of the ARPES autocorrelation and the QSI peaks in cuprate superconductors. Finally, we give a summary in Sec. \ref{conclusions}.

\section{Autocorrelation function of quasiparticle spectral intensities}\label{Formalism}

The ARPES autocorrelation of cuprate superconductors can be described in terms of the quasiparticle excitation spectrum as \cite{Chatterjee06},
\begin{eqnarray}\label{ACF}
{\bar C}({\bf q},\omega)&=&{1\over N}\sum_{\bf k}I({\bf k}+{\bf q},\omega)I({\bf k},\omega), ~~~~~~~
\end{eqnarray}
where the summation of momentum ${\bf k}$ is restricted within the first Brillouin zone (BZ), while the quasiparticle excitation spectrum $I({\bf k}, \omega)$ is related directly to the electron spectral function $A({\bf k},\omega)$ as $I({\bf k},\omega)=|M({\bf k},\omega)|^{2}n_{\rm F}(\omega) A({\bf k},\omega)$, with the fermion distribution $n_{\rm F}(\omega)$ and the dipole matrix element $M({\bf k},\omega)$. However, the important point is that $M({\bf k},\omega)$ does not have any significant energy or temperature dependence \cite{Zhou18,Damascelli03,Campuzano04,Kordyuk10}. In this case, the magnitude of $M({\bf k},\omega)$ can be rescaled to the unit, and then the evolution of ${\bar C}({\bf q},\omega)$ with momentum, energy, temperature, and doping concentration is completely characterized by the electron spectral function $A({\bf k},\omega)$. This autocorrelation function of the quasiparticle excitation spectral intensities ${\bar C}({\bf q},\omega)$ in Eq. (\ref{ACF}) therefore describes the ARPES autocorrelation at two different momenta, separated by a momentum transfer ${\bf q}$, at a fixed energy $\omega$.

The present discussion of the ARPES autocorrelation and its connection with QSI is based on the framework of the kinetic-energy-driven superconductivity \cite{Feng15,Feng0306,Feng12}. This kinetic-energy-driven SC mechanism is developed early within the $t$-$J$ model in the charge-spin separation fermion-spin representation, where the charge carriers are held together in the pairs in the particle-particle channel by the effective interaction that originates directly from the kinetic energy of the $t$-$J$ model by the exchange of spin excitations, then the electron pairs originating from the charge-carrier pairing state are due to the charge-spin recombination \cite{Feng15a}, and their condensation reveals the SC-state. In particular, this same interaction also generates the pseudogap state in the particle-hole channel \cite{Feng12}, leading to a coexistence of the SC-state and pseudogap state below the SC transition temperature $T_{\rm c}$ in the whole SC dome. Following these previous works, the single-particle diagonal and off-diagonal Green's functions $G({\bf k},\omega)$ and $\Im^{\dagger}({\bf k},\omega)$ of cuprate superconductors in the SC-state can be obtained as \cite{Feng15a},
\begin{widetext}
\begin{subequations}\label{EGF}
\begin{eqnarray}
G({\bf k},\omega)&=&{1\over \omega-\varepsilon_{\bf k}-\Sigma_{1}({\bf k},\omega)-[\Sigma_{2}({\bf k},\omega)]^{2}/[\omega+\varepsilon_{\bf k}+ \Sigma_{1}({\bf k},-\omega)]}, ~~~~~~ \label{DEGF}\\
\Im^{\dagger}({\bf k},\omega)&=&-{\Sigma_{2}({\bf k},\omega)\over [\omega-\varepsilon_{\bf k}-\Sigma_{1}({\bf k},\omega)][\omega+\varepsilon_{\bf k}+ \Sigma_{1}({\bf k},-\omega)]-[\Sigma_{2}({\bf k},\omega)]^{2}},~~~~\label{ODEGF}
\end{eqnarray}
\end{subequations}
\end{widetext}
where $\varepsilon_{\bf k}=-Zt\gamma_{\bf k}+Zt'\gamma_{\bf k}'+\mu$ is the bare band structure, with $\gamma_{\bf k}=({\rm cos}k_{x}+{\rm cos} k_{y})/2$, $\gamma_{\bf k}'= {\rm cos} k_{x}{\rm cos}k_{y}$, the nearest-neighbor (NN) and next NN electron hopping integrals $t$ and $t'$ in the $t$-$J$ model, respectively, the number of the NN or next NN sites on a square lattice $Z$ , and the chemical potential $\mu$. The electron self-energies $\Sigma_{1}({\bf k},\omega)$ in the particle-hole channel in Eq. (\ref{EGF}), which describes the single-particle coherence, and $\Sigma_{2}({\bf k},\omega)$ in the particle-particle channel, which is defined as the energy and momentum-dependence of the SC gap  \cite{Eliashberg60,Schrieffer64,Mahan81}, $\bar{\Delta}_{\rm s}({\bf k},\omega)=\Sigma_{2}({\bf k},\omega)$, can be obtained in terms of the full charge-spin recombination, and are given explicitly in Ref. \onlinecite{Feng15a}. In this paper, the parameters are chosen as $t/J=2.5$ and $t'/t=0.3$, where $J$ is the AF exchange in the $t$-$J$ model for a pair of NN spins. The magnitude of $J$ and the lattice constant of the square lattice are the energy and length units, respectively. However, when necessary to compare with the experimental data, we take $J=100$ meV \cite{Zhou18,Damascelli03,Campuzano04,Kordyuk10,Devereaux07,Fischer07}.

With the above single-particle diagonal Green's function (\ref{DEGF}), the electron spectral function $A({\bf k},\omega)$ in the SC-state now can be obtained explicitly as,
\begin{eqnarray}\label{ESF}
A({\bf k},\omega)={2\Gamma({\bf k},\omega)\over [\omega-E({\bf k},\omega)]^{2}+\Gamma^{2}({\bf k},\omega)},
\end{eqnarray}
where the quasiparticle scattering rate $\Gamma({\bf k},\omega)$ and the renormalized band structure $E({\bf k},\omega)$ are given by,
\begin{widetext}
\begin{subequations}\label{MESE}
\begin{eqnarray}
\Gamma({\bf k},\omega)&=&\left |{\rm Im}\Sigma_{1}({\bf k},\omega)-{\bar{\Delta}^{2}_{\rm s}({\bf k},\omega){\rm Im}\Sigma_{1}({\bf k},-\omega)\over [\omega +\varepsilon_{\bf k}+{\rm Re}\Sigma_{1}({\bf k}, -\omega)]^{2}+[{\rm Im}\Sigma_{1}({\bf k},-\omega)]^{2}}\right |, ~~~~~\label{EQDSR}\\
E({\bf k},\omega)&=&\varepsilon_{\bf k}+{\rm Re}\Sigma_{1}({\bf k},\omega) + {\bar{\Delta}^{2}_{\rm s}({\bf k},\omega)[\omega+\varepsilon_{\bf k}+ {\rm Re}\Sigma_{1}({\bf k},-\omega)]\over [\omega+\varepsilon_{\bf k}+{\rm Re}\Sigma_{1}({\bf k},-\omega)]^{2}+[{\rm Im}\Sigma_{1}({\bf k}, -\omega)]^{2}}, ~~~\label{MRESE}
\end{eqnarray}
\end{subequations}
\end{widetext}
with ${\rm Re}\Sigma_{1}({\bf k},\omega)$ and ${\rm Im}\Sigma_{1}({\bf k},\omega)$ that are the real and imaginary parts of the electron self-energy $\Sigma_{1}({\bf k},\omega)$, respectively. Substituting this electron spectral function $A({\bf k},\omega)$ into Eq. (\ref{ACF}), we therefore obtain the autocorrelation function of the quasiparticle excitation spectral intensities ${\bar C}({\bf q}, \omega)$.

\begin{figure*}[t!]
\centering
\includegraphics[scale=1.0]{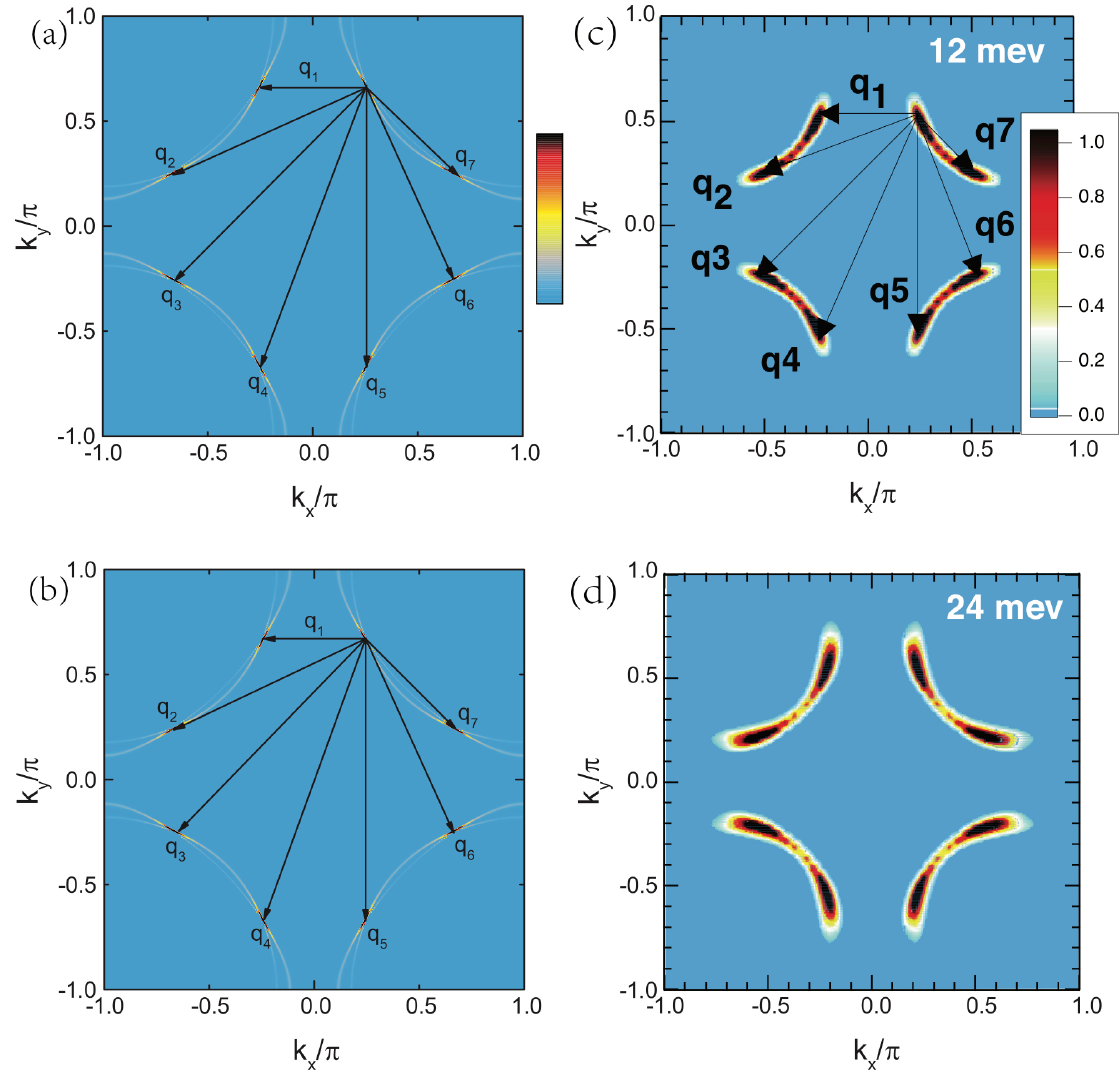}
\caption{(Color online) The spectral intensity of the quasiparticle excitation spectrum as a function of the momentum for (a) $\omega=0.12J=12$ meV and (b) $\omega=0.24J=24$ meV at $\delta=0.15$ with $T=0.002J$ for $t/J=2.5$, $t'/t=0.3$, and $J=100$ meV. The corresponding experimental results of the optimally doped Bi$_{2}$Sr$_{2}$CaCu$_{2}$O$_{8+\delta}$ for (c) $\omega=12$ meV and (d) $\omega=24$ meV taken from Ref. \onlinecite{Chatterjee06}.
\label{spectrum-maps}}
\end{figure*}

\section{ARPES autocorrelation and its connection with QSI} \label{ARPES-autocorrelation}

\begin{figure*}[t!]
\centering
\includegraphics[scale=1.0]{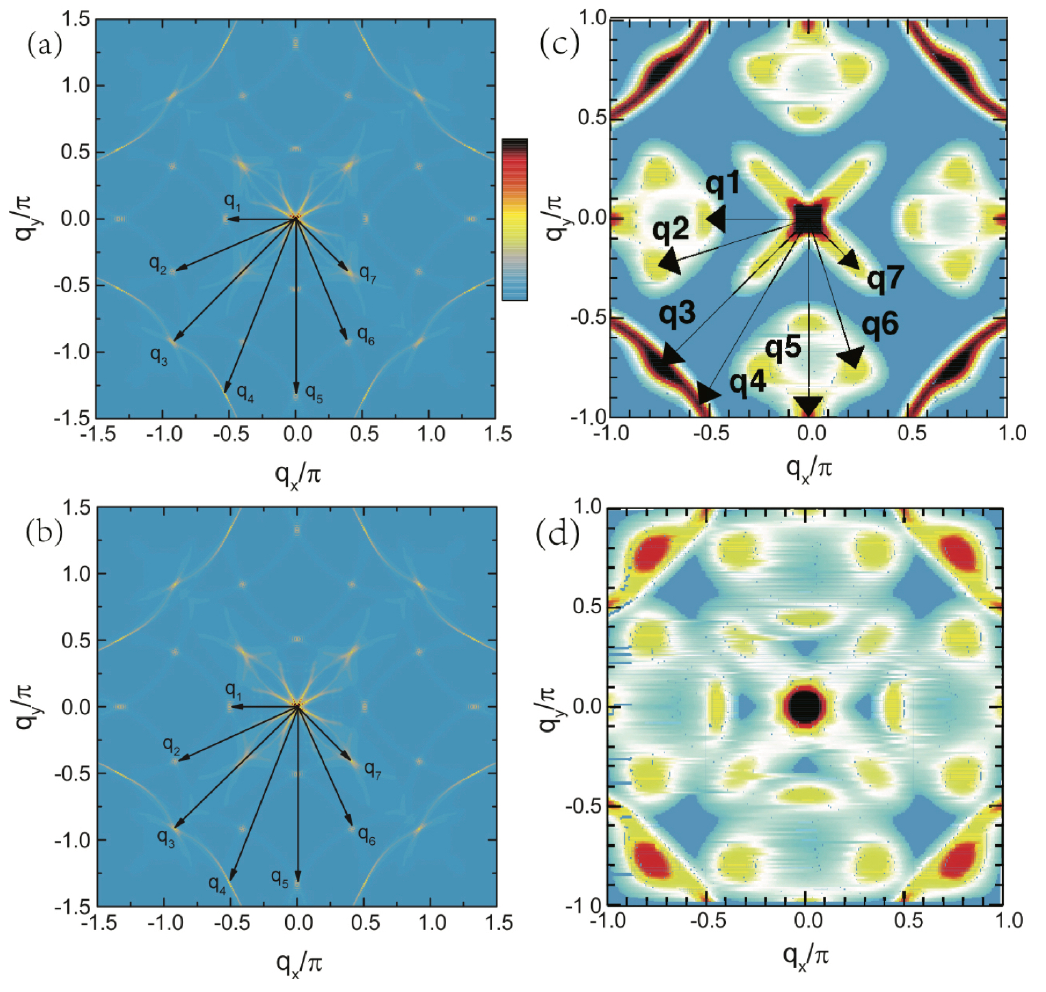}
\caption{(Color online) The autocorrelation of the quasiparticle excitation spectral intensities as a function of the momentum for (a) $\omega=0.12J= 12$ meV and (b) $\omega=0.24J=24$ meV at $\delta=0.15$ with $T=0.002J$ for $t/J=2.5$, $t'/t=0.3$, and $J=100$ meV. The corresponding experimental results of the optimally doped Bi$_{2}$Sr$_{2}$CaCu$_{2}$O$_{8+\delta}$ for (c) $\omega=12$ meV and (d) $\omega=24$ meV taken from Ref. \onlinecite{Chatterjee06}. \label{autocorrelation-maps}}
\end{figure*}

\begin{figure*}[t!]
\centering
\includegraphics[scale=1.0]{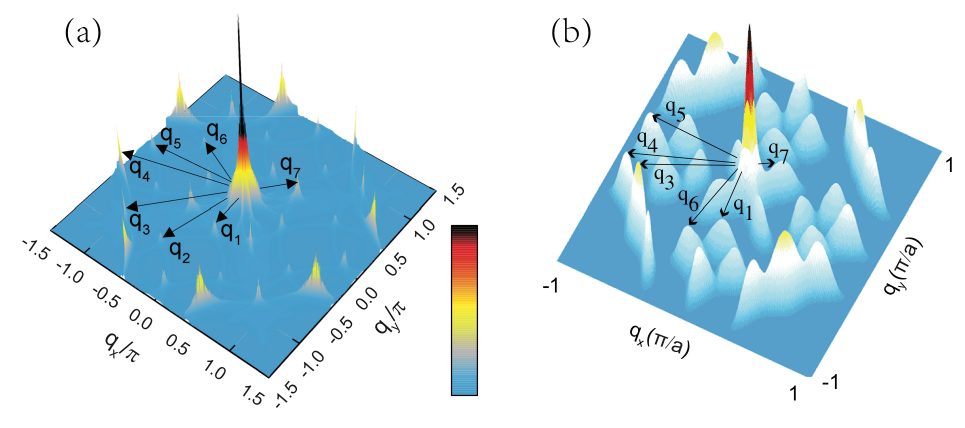}
\caption{(Color online) (a) The map of the intensity of the autocorrelation of the quasiparticle excitation spectral intensities in momentum-space for $\omega=0.18J=18$ meV at $\delta=0.15$ with $T=0.002J$ for $t/J=2.5$, $t'/t=0.3$, and $J=100$ meV. (b) The corresponding experimental result of the optimally doped Bi$_{2}$Sr$_{2}$CaCu$_{2}$O$_{8+\delta}$ for $\omega=18$ meV taken from Ref. \onlinecite{Chatterjee06}. \label{autocorrelation-peaks}}
\end{figure*}

In our recent studies \cite{Gao18a}, the nature of the SC-state EFS (the zero binding energy contour) reconstruction in cuprate superconductors has been discussed, where we have performed a map of the spectral intensity of the quasiparticle excitation spectrum $I({\bf k},0)$ at zero binding energy, and shown that the formation of the Fermi pockets due to the EFS reconstruction is closely related to the emergence of the highly anisotropic momentum-dependence of pseudogap. However, as a complement of these recent studies, we firstly in this paper discuss the nature of the SC-state constant energy contours at the case for finite binding energies. In Fig. \ref{spectrum-maps}, we plot a map of the spectral intensity of the SC-state quasiparticle excitation spectrum $I({\bf k},\omega)$ as a function of the momentum in the first BZ for the binding energies (a) $\omega= 0.12J=12$ meV and (b) $\omega=0.24J=24$ meV at the optimal doping $\delta=0.15$ with temperature $T=0.002J$. For comparison, the corresponding ARPES experimental results \cite{Chatterjee06} of the SC-state ARPES spectral intensity map observed on the optimally doped Bi$_{2}$Sr$_{2}$CaCu$_{2}$O$_{8+\delta}$ for the binding energies $\omega=12$ meV and $\omega=24$ meV are also shown in Fig. \ref{spectrum-maps}c and Fig. \ref{spectrum-maps}d, respectively. Apparently, the corresponding ARPES experimental results \cite{Chatterjee06} are qualitatively reproduced, where the most typical features are that: (i) the Fermi pockets formed by the disconnected segments on the constant energy contours that emerge due to the EFS reconstruction at the case for the zero binding energy \cite{Norman98,Yang08,Meng09,Yang11,Chan16} can persist into the case for finite binding energies. In particular, the area of these Fermi pockets is energy dependent; (ii) however, it is remarkable that the highest intensity points do not locate at the node places, but sit exactly at the tips of the disconnected segments, which in this case coincide with the hot spots on the constant energy contours, where we use the notation {\it hot spots} on the constant energy contours even for finite binding energies; (iii) since the most of the quasiparticles are accommodated at eight hot spots, these eight hot spots connected by the scattering wave vectors ${\bf q}_{i}$ shown in Fig. \ref{spectrum-maps}a therefore contribute effectively to the quasiparticle scattering processes \cite{Devereaux07,Fischer07,Pan01,Kohsaka07,Hanaguri07,Kohsaka08,Hanaguri09,Kondo09,Vishik09}. More specifically, these quasiparticle scattering processes with the scattering wave vectors ${\bf q}_{i}$ construct a {\it octet} scattering model \cite{Devereaux07,Fischer07,Pan01,Kohsaka07,Hanaguri07,Kohsaka08,Hanaguri09,Kondo09,Vishik09}. All these typical features are the same as the case for the zero binding energy \cite{Norman98,Yang08,Meng09,Yang11,Chan16} and are well consistent with the experimental observations \cite{Chatterjee06}. However, it should be emphasized that this microscopic {\it octet} scattering model with the scattering wave vectors ${\bf q}_{i}$ connecting the hot spots shown in Fig. \ref{spectrum-maps}a is obtained within the framework of the kinetic-energy-driven superconductivity. Furthermore, we \cite{Gao18a,Feng16} have shown that the quasiparticle scattering between two hot spots on the straight disconnected segments with the characteristic wave vector ${\bf q}_{1}={\bf Q}_{\rm HS}$ matches well with the corresponding charge-order wave vector ${\bf Q}_{\rm CD}$ observed in the resonant X-ray scattering measurements and STS experiments \cite{Comin16,Comin14,Wu11,Campi15,Comin15,Hinton16}.

\begin{figure*}[t!]
\centering
\includegraphics[scale=1.0]{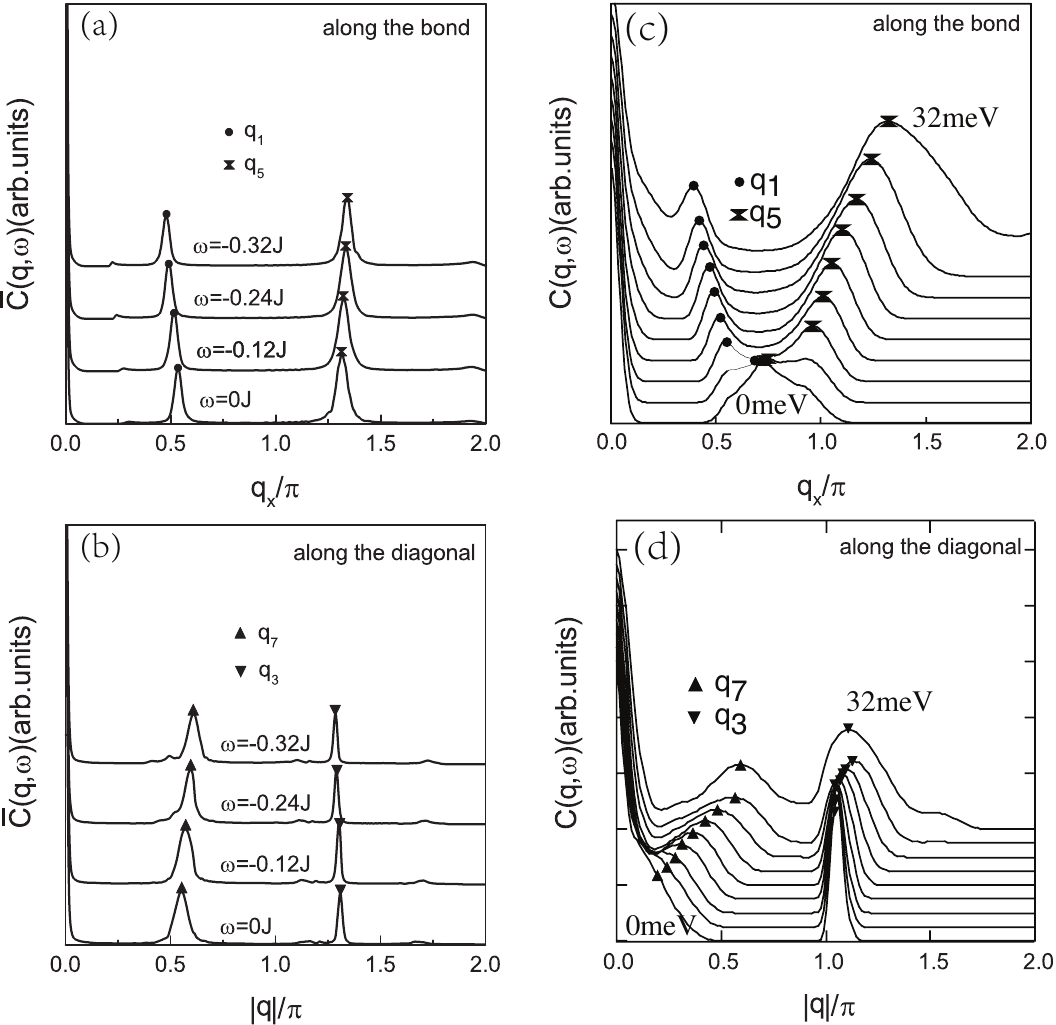}
\caption{The autocorrelation of the quasiparticle excitation spectral intensities as a function of the momentum along (a) the Brillouin zone parallel direction for the wave vectors ${\bf q}_{1}$ and ${\bf q}_{5}$ and (b) the Brillouin zone diagonal direction for the wave vectors ${\bf q}_{3}$ and ${\bf q}_{7}$ at $\delta=0.15$ with $T=0.002J$ for $t/J=2.5$ and $t'/t=0.3$. The corresponding experimental results of the ARPES autocorrelation of the optimally doped Bi$_{2}$Sr$_{2}$CaCu$_{2}$O$_{8+\delta}$ along (c) the Brillouin zone parallel direction for the wave vectors ${\bf q}_{1}$ and ${\bf q}_{5}$ and (d) the Brillouin zone diagonal direction for the wave vectors ${\bf q}_{3}$ and ${\bf q}_{7}$ taken from Ref. \onlinecite{Chatterjee06}. \label{autocorrelation-dispersion}}
\end{figure*}

We are now ready to discuss the ARPES autocorrelation and its connection with QSI in cuprate superconductors. In Fig. \ref{autocorrelation-maps}, we plot the autocorrelation of the SC-state quasiparticle excitation spectral intensities ${\bar C}({\bf q},\omega)$ as a function of the momentum for the binding energies (a) $\omega=0.12J=12$ meV and (b) $\omega=0.24J=24$ meV at $\delta=0.15$ with $T=0.002J$ in comparison with the corresponding experimental results \cite{Chatterjee06} of the ARPES autocorrelation detected from the optimally doped Bi$_{2}$Sr$_{2}$CaCu$_{2}$O$_{8+\delta}$ for the bind energies (c) $\omega=12$ meV and (d) $\omega=24$ meV, respectively. Our results thus show that there are some sharp peaks (then some discrete spots) appear in ${\bar C}({\bf q},\omega)$, where the joint density of states is highest. To see these sharp peaks at the discrete spots of ${\bar C}({\bf q},\omega)$ more clearly, we map ${\bar C}({\bf q},\omega)$ in the $[k_{x},k_{y}]$ plane for the binding energy $\omega=0.18J=18$ meV at $\delta=0.15$ with $T=0.002J$ in Fig. \ref{autocorrelation-peaks}a in comparison with the corresponding experimental result \cite{Chatterjee06} of the ARPES autocorrelation of the optimally doped Bi$_{2}$Sr$_{2}$CaCu$_{2}$O$_{8+\delta}$ for the binding energy $\omega=18$ meV in Fig. \ref{autocorrelation-peaks}b, where the sharp autocorrelation peaks are located exactly at the discrete spots of ${\bar C}({\bf q},\omega)$. Moreover, these combined results in Fig. \ref{autocorrelation-maps} and Fig. \ref{autocorrelation-peaks} therefore indicate clearly that the discrete spots in ${\bar C}({\bf q},\omega)$ are directly correlated with those wave vectors ${\bf q}_{i}$ connecting the hot spots on the constant energy contours shown in Fig. \ref{spectrum-maps}a, in good agreement with the ARPES experimental data \cite{Chatterjee06}. For a further understanding of the anomalous properties of the energy- and momentum-dependence of ${\bar C} ({\bf q},\omega)$, we have made a series of calculations for ${\bar C} ({\bf q},\omega)$ with the different ${\bf q}_{i}$, and the results of ${\bar C}({\bf q},\omega)$ as a function of the momentum along (a) the BZ parallel direction for the scattering wave vectors ${\bf q}_{1}$ and ${\bf q}_{5}$ and (b) the BZ diagonal direction for the scattering wave vectors ${\bf q}_{3}$ and ${\bf q}_{7}$ at $\delta=0.15$ with $T=0.002J$ are plotted in Fig. \ref{autocorrelation-dispersion}. For comparison, the corresponding APRES experimental results \cite{Chatterjee06} of the optimally doped Bi$_{2}$Sr$_{2}$CaCu$_{2}$O$_{8+\delta}$ along the BZ parallel direction for the scattering wave vectors ${\bf q}_{1}$ and ${\bf q}_{5}$ and the BZ diagonal direction for the scattering wave vectors ${\bf q}_{3}$ and ${\bf q}_{7}$ are also shown in Fig. \ref{autocorrelation-dispersion}c and Fig. \ref{autocorrelation-dispersion}d, respectively. These results therefore show that the sharp peaks in ${\bar C}({\bf q},\omega)$ at low energies disperse smoothly with energy. In particular, the dispersion of the sharp peaks in ${\bar C}({\bf q},\omega)$ follow the evolution of the hot spots on the disconnected segments with energy, and are also qualitatively consistent with the ARPES observations \cite{Chatterjee06}.

\begin{figure*}[t!]
\centering
\includegraphics[scale=1.0]{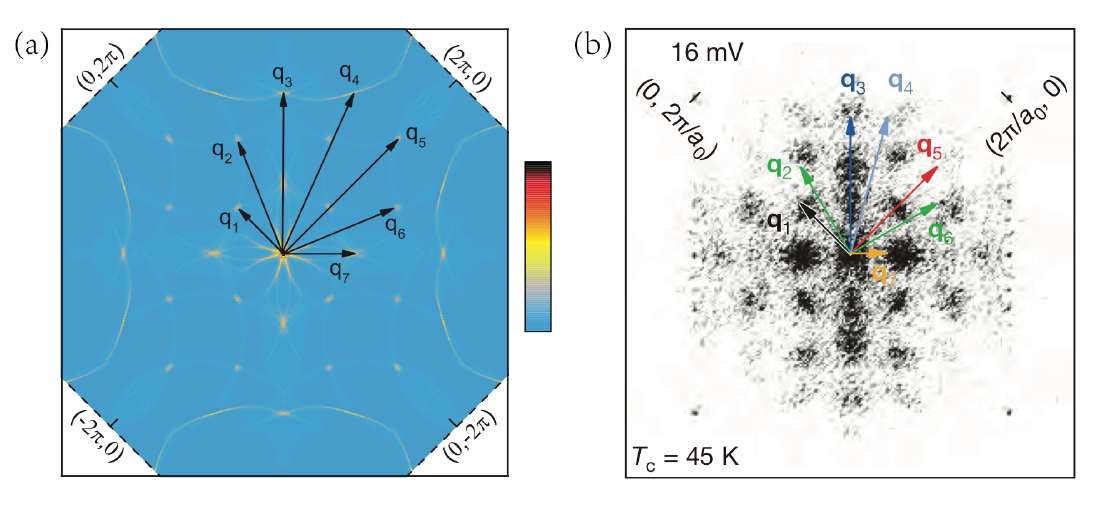}
\caption{(Color online) (a) The autocorrelation of the quasiparticle excitation spectral intensities as a function of the momentum for $\omega=0.16J= 16$ meV at $\delta=0.15$ with $T=0.002J$ for $t/J=2.5$, $t'/t=0.3$, and $J=100$ meV. (b) The experimental result of the quasiparticle scattering interference pattern $Z({\bf q},\omega=16$ meV) for Bi$_{2}$Sr$_{2}$CaCu$_{2}$O$_{8+\delta}$ taken from Ref. \onlinecite{Kohsaka08}. \label{autocorrelation-octet}}
\end{figure*}

On the other hand, it is very remarkable that the momentum-space structure of the ARPES autocorrelation patterns connected by the scattering wave vectors ${\bf q}_{i}$ is well consistent with the momentum-space structure of the QSI patterns observed from the FT-STS experiments \cite{Devereaux07,Fischer07,Pan01,Kohsaka07,Hanaguri07,Kohsaka08,Hanaguri09,Kondo09,Vishik09}. In Fig. \ref{autocorrelation-octet}a, we plot ${\bar C}({\bf q},\omega)$ as a function of the momentum for the binding energy $\omega=0.16J=16$ meV at $\delta=0.15$ with $T=0.002J$. For comparison, the experimental result \cite{Kohsaka08} of the QSI pattern obtained from the optimally doped Bi$_{2}$Sr$_{2}$CaCu$_{2}$O$_{8+\delta}$ for the binding energy $\omega=16$ meV is also shown in Fig. \ref{autocorrelation-octet}b. It is clear that the momentum-space structure of the ARPES autocorrelation pattern shown in Fig. \ref{autocorrelation-octet}a is in good agreement with the momentum-space structure of the QSI pattern shown in Fig. \ref{autocorrelation-octet}b. Moreover, combining the results in Fig. \ref{spectrum-maps}, Fig. \ref{autocorrelation-maps}, and Fig. \ref{autocorrelation-octet} thus confirm that the {\it octet} scattering model constructed by the eight hot spots shown in Fig. \ref{spectrum-maps}a that can give a consistent description of the regions of the highest joint density of states can be also used to explain the FT-STS experimental data.

\begin{figure*}[t!]
\centering
\includegraphics[scale=1.0]{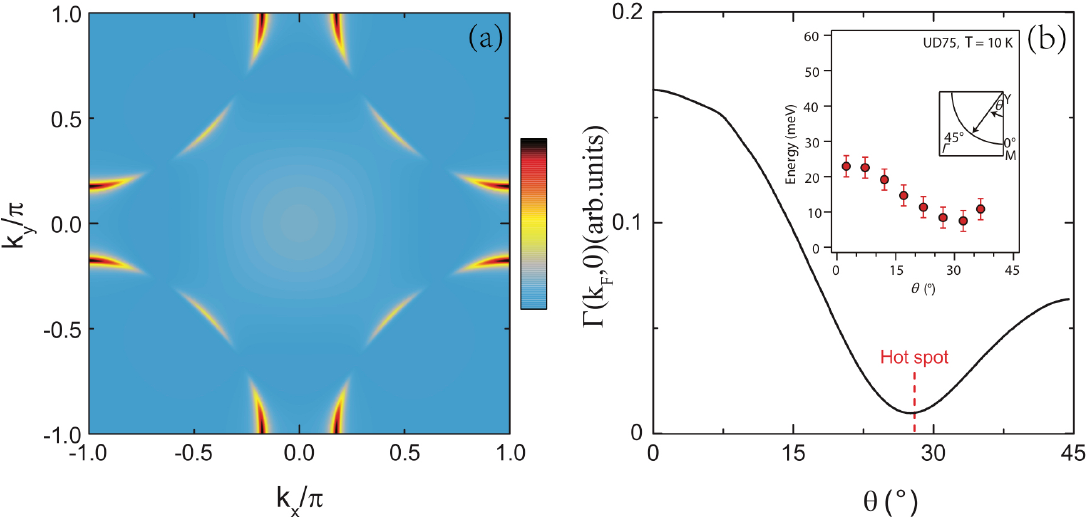}
\caption{(Color online) (a) The map of the intensity of the quasiparticle scattering rate and (b) the angular dependence of the the quasiparticle scattering rate on the constant energy contour shown in Fig. \ref{spectrum-maps}b in $\omega=0.24J=24$ meV at $\delta=0.09$ with $T=0.002J$ for $t/J=2.5$, $t'/t=0.3$, and $J=100$ meV.  Inset in (b): the corresponding experimental result of the angular dependence of the quasiparticle scattering rate for the underdoped Bi$_{2}$Sr$_{2}$CaCu$_{2}$O$_{8+\delta}$ taken from Ref. \onlinecite{Vishik09}. \label{scattering-rate}}
\end{figure*}

In our previous studies \cite{Gao18a,Feng16}, we have shown clearly that the formation of the Fermi pockets at the zero binding energy and the related striking feature of the sharp quasiparticle peak with the large spectral weight appeared always at the hot spots on EFS can be attributed to the highly anisotropic momentum-dependence of the quasiparticle scattering rate $\Gamma({\bf k},0)$ in Eq. (\ref{EQDSR}) at the zero binding energy. Our present results therefore show that the essential physics of the formation of the Fermi pockets at finite binding energies and the related the {\it octet} scattering model with the scattering wave vectors ${\bf q}_{i}$ connecting the hot spots shown in Fig. \ref{spectrum-maps}a is the same as that of the zero binding energy, and can be also attributed to the highly anisotropic momentum-dependence of the quasiparticle scattering rate $\Gamma({\bf k},\omega)$ at finite binding energies. In Fig. \ref{scattering-rate}, we plot (a) the map of the intensity of $\Gamma({\bf k},\omega)$ in the first BZ and (b) the angular dependence of $\Gamma({\bf k},\omega)$ on the constant energy contour shown in Fig. \ref{spectrum-maps}b for the binding energy $\omega=0.24J=24$ meV at $\delta=0.09$ with $T=0.002J$ in comparison with the corresponding experimental result \cite{Vishik09} of the angular dependence of the quasiparticle scattering rate along the constant energy contour observed on the underdoped Bi$_{2}$Sr$_{2}$CaCu$_{2}$O$_{8+\delta}$ (inset in b), where as in the case of the zero binding energy \cite{Gao18a,Feng16}, the actual minimum of $\Gamma({\bf k},\omega)$ appears always at the off-node place, and therefore is located exactly at the hot spot on EFS. However, the largest value of $\Gamma({\bf k},\omega)$ still emerges at the antinode, and then it damps down when the momentum shifts away from the antinode. In particular, the magnitude of $\Gamma({\bf k},\omega)$ around the antinode is always larger than that around the node. This highly anisotropic momentum-dependence of $\Gamma({\bf k},\omega)$ thus reduces heavily the spectral weight of the electron quasiparticle excitation spectrum around the antinodal region, but it has a more modest effect on the spectral weight around the nodal region, and then the tips of these disconnected segments on the constant energy contours converge on the eight hot spots to form the closed Fermi pockets. These eight hot spots construct a {\it octet} scattering model with the scattering wave vectors ${\bf q}_{i}$ as shown in Fig. \ref{spectrum-maps}b, which therefore leads to that the sharp peaks in the ARPES autocorrelation with the scattering wave vectors ${\bf q}_{i}$ are directly correlated to the regions of the highest joint density of states.

\begin{figure*}[t!]
\centering
\includegraphics[scale=1.0]{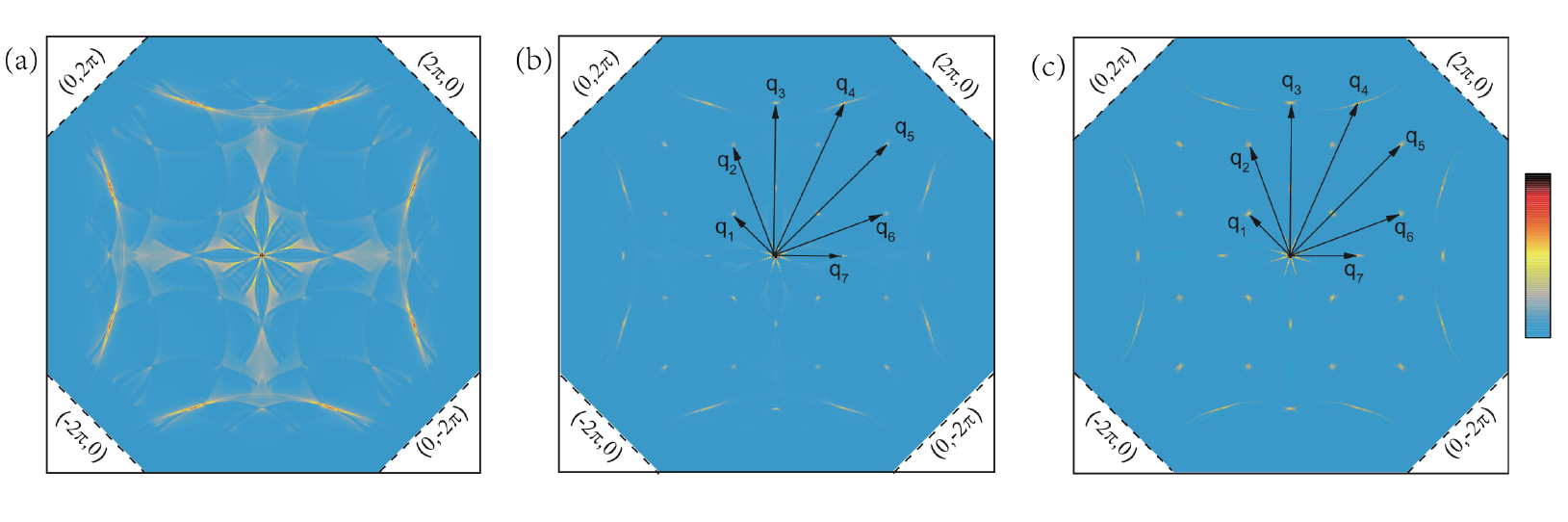}
\caption{(Color online) The Fourier transformed local density of states as a function of the momentum for $\omega=0.16J=16$ meV at $\delta=0.15$ with $T=0.002J$ for $t/J=2.5$, $t'/t=0.3$, and $J=100$ meV in the presence of single point-like potential scatterer of strength (a) $V_{0}=0.1J=10$ meV, (b) $V_{0}=0.5J=50$ meV, and (c) $V_{0}=1.0J=100$ meV. \label{QSI-LDOS}}
\end{figure*}

However, this momentum-dependence of $\Gamma({\bf k},\omega)$ at finite binding energies is also intrinsically related to the emergence of the momentum-dependence of the pseudogap, since the electron self-energy $\Sigma_{1}({\bf k},\omega)$ in the particle-hole channel in Eq. (\ref{EGF}) can be also rewritten as $\Sigma_{1}({\bf k},\omega)\approx [\bar{\Delta}_{\rm PG}({\bf k})]^{2}/(\omega+ \varepsilon_{0{\bf k}})$ with the energy spectrum $\varepsilon_{0{\bf k}}$ and the pseudogap $\bar{\Delta}_{\rm PG}({\bf k})$ that have been given explicitly in Ref. \onlinecite{Feng15a}. Moreover, the imaginary part of $\Sigma_{1}({\bf k},\omega)$ in the quasiparticle scattering rate (\ref{EQDSR}) is closely related to the pseudogap as ${\rm Im}\Sigma_{1}({\bf k},\omega)\approx 2\pi[\bar{\Delta}_{\rm PG}({\bf k})]^{2}\delta(\omega+\varepsilon_{0{\bf k}})$, reflecting a fact that this pseudogap $\bar{\Delta}_{\rm PG}({\bf k})$ has the same angular dependence on EFS as that of $\Gamma({\bf k},\omega)$. This connection of the pseudogap $\bar{\Delta}_{\rm PG}({\bf k})$ and the quasiparticle scattering rate $\Gamma({\bf k},\omega)$ therefore shows that the formation of the Fermi pockets and the related the {\it octet} scattering structure with the scattering wave vectors ${\bf q}_{i}$ connecting the hot spots shown in Fig. \ref{spectrum-maps} are intrinsically associated to the emergence of the momentum-dependence of the pseudogap.

We now turn to show why there is an intimate connection between the ARPES autocorrelation and QSI in cuprate superconductors. Theoretically, the QSI experiments have been interpreted in terms of the phenomenological {\it octet} scattering model by considering the scattering arising from a single point-like impurity \cite{Devereaux07,Fischer07,Wang03,Wang10,Nunner06,Zhang03}. Within the framework of the kinetic-energy-driven SC mechanism, the inhomogeneous part of $\rho({\bf q},\omega)$ in the presence of a single point-like impurity scattering potential $\tilde{V}=V_{0}\delta({\bf r}) \tau_{3}$ can be evaluated in terms of the electron diagonal and off-diagonal Green's functions $G({\bf k},\omega)$ and $\Im^{\dagger}({\bf k}, \omega)$ in Eq. (\ref{EGF}) as,
\begin{eqnarray}\label{LDOS}
\delta\rho({\bf q},\omega)&=&-{1\over\pi}{\rm Im}{1\over N}\sum_{\bf k}[G({\bf k}+{\bf q},\omega)T_{1}(\omega)G({\bf k},\omega)\nonumber\\
&+&\Im({\bf k}+{\bf q}, \omega) T_{2}(\omega)\Im^{\dagger}({\bf k},\omega)],~~~~
\end{eqnarray}
where $T_{1}(\omega)$ and $T_{2}(\omega)$ are the energy-dependent elements of the $\tilde{T}$ matrix, and can be expressed explicitly as,
\begin{subequations}\label{T-matrix}
\begin{eqnarray}
T_{1}(\omega)&=&{V_{0}\over [1-G(\omega)V_{0}]}, \\
T_{2}(\omega)&=&-{V_{0}\over [1-G(-\omega)V_{0}]},
\end{eqnarray}
\end{subequations}
with $G(\omega)=(1/N)\sum_{{\bf k}}G({\bf k},\omega)$.

In Fig. \ref{QSI-LDOS}, we plot the momentum-space patterns of $\delta\rho({\bf q},\omega)$ for the binding energy $\omega=0.16J=16$ meV at $\delta=0.15$ with $T=0.002J$ in the presence of a single point-like potential scatterer of the strengths (a) $V_{0}=0.1J=10$ meV, (b) $V_{0}=0.5J=50$ meV, and (c) $V_{0}=1.0J=100$ meV, where the single point-like potential scatterer of the strength $V_{0}<0.5J$ can be thought to be a scattering process with the weak scattering potential, while $V_{0}>0.5J$ is a scattering process with the strong scattering potential. In particular, our results show that for the scattering process with the strong scattering potential ($V_{0}>0.5J$), the obtained momentum-space structure of the $\delta\rho({\bf q},\omega)$ patterns in the SC-state is qualitatively consistent with the momentum-space structure of the QSI patterns \cite{Devereaux07,Fischer07,Pan01,Kohsaka07,Hanaguri07,Kohsaka08,Hanaguri09,Kondo09,Vishik09} observed from the FT-STS experiments on cuprate superconductors in the SC-state. However, it is very surprising that the obtained result of the momentum-space structure of the $\delta\rho({\bf q},\omega)$ patterns in the strong scattering process is also in qualitative agreement with the momentum-space structure of the ARPES autocorrelation patterns shown in Fig. \ref{autocorrelation-maps}c and Fig. \ref{autocorrelation-maps}d found from the ARPES measurements \cite{Chatterjee06}.

The equation (\ref{LDOS}) indicates that there are two parts of the contribution to $\delta\rho({\bf q},\omega)$: the contribution from the first term of the right-hand side in Eq. (\ref{LDOS}) comes from the quasiparticle scattering process in the presence of a single point-like impurity in the particle-hole channel obtained in terms of the electron diagonal Green's function, and therefore is closely associated with the pseudogap $\bar{\Delta}_{\rm PG}$, while the additional contribution from the second term of the right-hand side in Eq. (\ref{LDOS}) originates from the quasiparticle scattering process in the presence of a single point-like impurity in the particle-particle channel obtained in terms of the electron off-diagonal Green's function, and therefore is directly connected with the SC gap $\bar{\Delta}_{\rm s}$. The strength of this additional quasiparticle scattering process in the particle-particle channel is proportional to $\bar{\Delta}^{2}_{\rm s}$. However, in the underdoped and optimally doped regimes, $\bar{\Delta}_{\rm s}\ll \bar{\Delta}_{\rm PG}$, and therefore the strength of the additional quasiparticle scattering process in the particle-particle channel is much smaller than that of the quasiparticle scattering process in the particle-hole channel. In other words, the contribution to $\delta\rho({\bf q},\omega)$ in the underdoped and optimally doped regimes is mainly dominated by the quasiparticle scattering process in the particle-hole channel. In particular, we find during the calculations that in the case of the presence of the strong scattering potential $V_{0}>0.5J$, the main contribution to $\delta\rho({\bf q},\omega)$ comes from the term ${\rm Im}G({\bf k}+{\bf q},\omega){\rm Im} T_{1}(\omega){\rm Im}G({\bf k},\omega)$ of the right-hand side in Eq. (\ref{LDOS}), and then $\delta\rho({\bf q},\omega)$ in Eq. (\ref{LDOS}) can be reduced as,
\begin{eqnarray}\label{LDOS-2}
\delta\rho({\bf q},\omega)&\approx& {1\over\pi}{1\over N}\sum_{\bf k}{\rm Im}G({\bf k}+{\bf q},\omega){\rm Im}T_{1}(\omega){\rm Im}G({\bf k},\omega) \nonumber\\
&\propto& {1\over N}\sum_{\bf k}A({\bf k}+{\bf q},\omega)A({\bf k},\omega),
\end{eqnarray}
where ${\rm Im}G({\bf k},\omega)$ and ${\rm Im}T_{1}(\omega)$ are the corresponding imaginary parts of $G({\bf k},\omega)$ and $T_{1}(\omega)$, respectively. This expression form in Eq. (\ref{LDOS-2}) is the same as the autocorrelation function of the quasiparticle excitation spectral intensities in Eq. (\ref{ACF}). This is why in the case of the presence of the strong scattering potential $V_{0}>0.5J$, the momentum-space structure of the QSI patterns \cite{Devereaux07,Fischer07,Pan01,Kohsaka07,Hanaguri07,Kohsaka08,Hanaguri09,Kondo09,Vishik09} is qualitatively consistent with the momentum-space structure of the ARPES autocorrelation patterns \cite{Chatterjee06} shown in Fig. \ref{autocorrelation-maps}c and Fig. \ref{autocorrelation-maps}d. The qualitative agreement between the momentum-space structure of the ${\bar C}({\bf q},\omega)$ patterns and the momentum-space structure of the $\delta\rho({\bf q},\omega)$ patterns in the case of the presence of the strong scattering potential therefore confirm an intimate connection between the ARPES autocorrelation and QSI in cuprate superconductors.

\section{Conclusions}\label{conclusions}

In summary, within the framework of the kinetic-energy driven SC mechanism, we have discussed the ARPES autocorrelation and its connection with QSI in cuprate superconductors by taking into account the pseudogap effect. Our results show that the {\it octet} scattering model of the quasiparticle scattering processes with the scattering wave vectors ${\bf q}_{i}$ connecting the hot spots on the constant energy contours is intrinsically related to the emergence of the highly anisotropic momentum-dependence of the pseudogap. This {\it octet} scattering model therefore leads to that the sharp peaks in the ARPES autocorrelation with the scattering wave vectors ${\bf q}_{i}$ are directly correlated to the regions of the highest joint density of states. Concomitantly, the momentum-space structure of the ARPES autocorrelation patterns detected from the ARPES experimental measurements is qualitatively consistent with the momentum-space structure of the QSI patterns observed from FT-STS experiments. Our theory therefore also confirms a direct connection between the ARPES autocorrelation and QSI in cuprate superconductors.

\section*{Acknowledgements}

The authors would like to thank Dr. Huaisong Zhao for the helpful discussions. This work was supported by the National Key Research and Development Program of China under Grant No. 2016YFA0300304, and the National Natural Science Foundation of China under Grant Nos. 11574032 and 11734002.

\end{document}